\documentclass[prl,superscriptaddress,showpacs,twocolumn,amsmath,amssymb]{revtex4-1}

\usepackage[dvips]{graphicx}
\usepackage{graphicx}% Include figure files
\usepackage{color}

\begin{document}

Published in: PRB \textbf{93}, 121202(R) (March 2016)

\title{Interband optical conductivity of the [001]-oriented Dirac semimetal Cd$_{3}$As$_{2}$}

\author{D. Neubauer}
\affiliation{1. Physikalisches Institut, Universit\"{a}t Stuttgart,
Pfaffenwaldring 57, 70550 Stuttgart, Germany}
\author{J. P. Carbotte}
\affiliation{Department of Physics and Astronomy, McMaster
University, Hamilton, Ontario L8S 4M1, Canada} \affiliation{The
Canadian Institute for Advanced Research, Toronto, Ontario M5G 1Z8,
Canada}
\author{A. A. Nateprov}
\affiliation{Institute of Applied Physics, Academy of Sciences of
Moldova, Academiei str. 5, 2028 Chisinau, Moldova}
\author{A. L\"{o}hle}
\affiliation{1. Physikalisches Institut, Universit\"{a}t Stuttgart,
Pfaffenwaldring 57, 70550 Stuttgart, Germany}
\author{M. Dressel}
\affiliation{1. Physikalisches Institut, Universit\"{a}t Stuttgart,
Pfaffenwaldring 57, 70550 Stuttgart, Germany}
\author{A. V. Pronin}
\email{artem.pronin@pi1.physik.uni-stuttgart.de} \affiliation{1.
Physikalisches Institut, Universit\"{a}t Stuttgart, Pfaffenwaldring
57, 70550 Stuttgart, Germany}

\date{13 January 2016}

\begin{abstract}
We measured the optical reflectivity of [001]-oriented $n$-doped
Cd$_{3}$As$_{2}$ in a broad frequency range (50 --
22\,000~cm$^{-1}$) for temperatures from 10 to 300~K. The optical
conductivity, $\sigma(\omega) = \sigma_{1}(\omega) + {\rm
i}\sigma_{2}(\omega)$, is isotropic within the (001) plane; its real
part follows a power law, $\sigma_{1}(\omega) \propto
\omega^{1.65}$, in a large interval from 2000 to 8000~cm$^{-1}$.
This behavior is caused by interband transitions between two Dirac
bands, which are effectively described by a sublinear dispersion
relation, $E(k) \propto \lvert k \rvert ^{0.6}$. The
momentum-averaged Fermi velocity of the carriers in these bands is
energy dependent and ranges from $1.2\times 10^{5}$ to $3 \times
10^{5}$~m/s, depending on the distance from the Dirac points. We
detect a gaplike feature in $\sigma_{1}(\omega)$ and associate it
with the Fermi level positioned around $100$~meV above the Dirac
points.\\
\\
DOI: 10.1103/PhysRevB.93.121202

\end{abstract}

\maketitle

\section{Introduction}

The interest in measurements of the optical conductivity,
$\sigma(\omega) = \sigma_{1}(\omega) + {\rm i}\sigma_{2}(\omega)$,
in three-dimensional (3D) Dirac materials \cite{Wehling} is
triggered by the fact that the interband conductivity in these
systems is expected to demonstrate a peculiar behavior. Generally,
the interband optical response of $d$-dimensional Dirac electrons is
supposed to be universal: $\sigma_{1}(\omega)$ should follow a
power-law frequency dependence,
\begin{equation}
\sigma_{1}(\omega) \propto \omega ^{(d - 2)/z}, \label{exponent}
\end{equation}
where $z$ is the exponent in the band dispersion relation, $E(k)
\propto \lvert k \rvert ^{z}$, \cite{Hosur2012, Bacsi2013}.

For example, $\sigma_{1}$ is proportional to frequency in the case
of perfectly linear Dirac cones in three dimensions. Such linearity
in $\sigma_{1}(\omega)$ over a broad frequency range in a 3D system
is often considered as a ``smoking gun'' for Dirac physics (either
of topological or of other origin, \cite{Orlita2014}). For example,
Timusk \textit{et al.} \cite{Timusk2013} have claimed the presence
of 3D Dirac fermions in a number of quasicrystals based entirely on
the observation of a linear $\sigma_{1}(\omega)$ in these materials.
Linear-in-frequency $\sigma_{1}(\omega)$ has also been found in
ZrTe$_{5}$ \cite{Chen2015}, where 3D linear bands are evidenced by
transport and angle-resolved photoemission experiments
\cite{Li2014}.

\begin{figure}[t]
\centering
\includegraphics[width=0.9\columnwidth,clip]{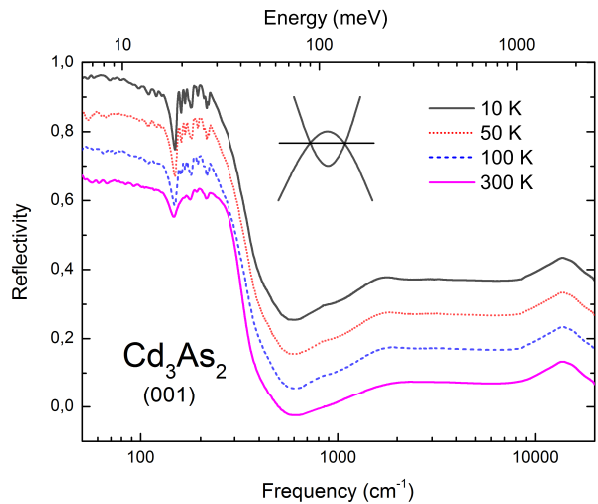}
\caption{(Color online) Reflectivity of [001]-cut Cd$_{3}$As$_{2}$
for selected temperatures between 10 and 300~K measured with
$\mathbf{E}_{\omega} \parallel$ [110], see text. Ordinate numbers
are given for the measurements at 10~K, while the curves obtained at
$T= 50$, $100$, and $300$ K are downshifted by $-0.1$, $-0.2$, and
$-0.3$, respectively. The inset sketches the dispersion of the Dirac
bands in Cd$_{3}$As$_{2}$. In general, the Fermi-level position is
not necessarily at the Dirac points.} \label{reflectivity}
\end{figure}

In 2013 Wang \textit{et al.} \cite{Wang2013} predicted 3D
topological Dirac points in Cd$_{3}$As$_{2}$; by now they are well
confirmed by ARPES, scanning tunneling spectroscopy, and
magnetotransport measurements \cite{Borisenko2014, Liu2014Nat,
Neupane2014, Jeon2014, He2014, Liang2015}. Due to the presence of
inversion symmetry in Cd$_{3}$As$_{2}$, the bands are not spin
polarized \cite{Ali2014}. The shape of the Dirac bands in
Cd$_{3}$As$_{2}$ is somewhat complicated by the presence of a
Lifshitz-transition point \cite{Jeon2014}; see
Fig.~\ref{reflectivity} for a sketch of the Dirac bands.

The goal of this report is to provide insight into the Dirac-band
dispersion in Cd$_{3}$As$_{2}$ by means of optical spectroscopy.
Previous optical investigations of Cd$_{3}$As$_{2}$ performed in the
70s and 80s can be divided into two groups. The first one
\cite{Turner1961, Iwami1973, Gelten1980, GeltenPhD, Houde1986, LB}
deals with the low-energy part of the optical spectra (usually,
below some 250~meV, or 2000~cm$^{-1}$) discussing mostly phonon
modes and free-electron Drude-like absorption. Turner \textit{et
al.} \cite{Turner1961} identify a very narrow (130~meV) optical gap
in the optical absorption. Another group of papers \cite{Zivitz,
Karnicka, Sobolev1, Sobolev2, Kozlov} mainly discusses absorption
features at energies of a few electron volts and their relations to
transitions between different (high-energy) parabolic bands. No
optical conductivity was derived from these measurements.

Recent recognition of the nontrivial electron-band topology of
Cd$_{3}$As$_{2}$ calls for a fresh look into its optical properties.
In this paper, we report broadband optical investigations of
Cd$_{3}$As$_{2}$.

\section{Experiment}

Single crystals of Cd$_{3}$As$_{2}$ have been grown by vapor
transport from material previously synthesized in argon flow
\cite{Nateprov2015}, see Supplemental Material for details.
Resistivity and Hall measurements provide an electron density of
$n_{e} = 6 \times 10^{17}$~cm$^{-3}$ (roughly independent of
temperature), a metallic resistivity, and a mobility of $\mu =
8\times 10^{4}$~cm$^{2}$/Vs at 12~K.

The investigated Cd$_{3}$As$_{2}$ single crystal had lateral
dimensions of 2.5~mm by 3~mm and a thickness of 300~$\mu$m. It was
cut out from a larger single crystal. The crystallographic axes of
the sample were found by x-ray diffraction. The [001] axis was
perpendicular to the sample's largest surface. This surface was
polished prior the optical measurements, which were performed for a
few linear polarizations. The direct-current (dc) resistivity of
this sample was characterized in-plane by standard four-probe method
(inset of Fig.~\ref{low_freq}).

The optical reflectivity was measured at 10 to 300~K with light
polarized along different crystallographic directions. The spectra
in the far-infrared (50~cm$^{-1}$ -- 1000~cm$^{-1}$) were recorded
by a Bruker IFS 113v Fourier-transform infrared spectrometer using
an in-situ gold overfilling technique for reference measurements
\cite{gold}. At higher frequencies (700~cm$^{-1}$ -- 22\,000
cm$^{-1}$) a Bruker Hyperion microscope attached to a Bruker Vertex
80v spectrometer was used. Here, either freshly evaporated gold
mirrors or coated silver mirrors were utilized as references.

Both dc and optical measurements revealed an isotropic response
within the (001)-plane. Hereafter, we present the optical data
obtained for $\mathbf{E}_{\omega} \parallel$ [110] where
$\mathbf{E}_{\omega}$ is the electric-field component of the probing
light.

\section{Experimental Results}

Figure~\ref{reflectivity} shows the reflectivity $R(\omega)$ of
[001]-oriented Cd$_{3}$As$_{2}$ versus frequency at various
temperatures as indicated. At low frequencies the reflectivity is
rather high corresponding to the metallic behavior of the dc
resistivity. At $\omega^{\rm scr}_{\rm pl}/2\pi c\approx
400$~cm$^{-1}$ a temperature-independent (screened) plasma edge is
observed in the reflectivity. A number of phonon modes strongly
affect the reflectivity at lower $\omega$. For $\omega/2\pi c
> 2000$~cm$^{-1}$ the reflectivity is basically temperature
independent.

\begin{figure}[b]
\centering
\includegraphics[width=0.9\columnwidth, clip]{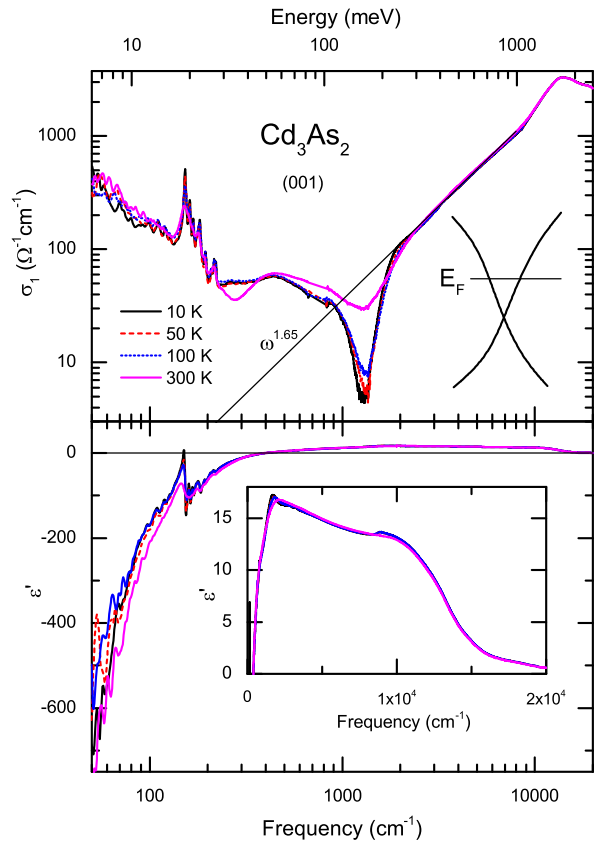}
\caption{(Color online) Overall optical conductivity on log-log
scale (upper frame) and dielectric permittivity (bottom frame) of
Cd$_{3}$As$_{2}$ measured within the (001) plane. The straight line
in the upper frame represents $\sigma_{1}(\omega) \propto
\omega^{1.65}$. The diagram in the upper frame sketches the proposed
band dispersion in Cd$_{3}$As$_{2}$ near one of the Dirac points.
The inset of the bottom frame displays positive
$\varepsilon^{\prime}$ on a linear $x$ scale.} \label{overall}
\end{figure}

In order to extract the complex optical conductivity, we applied a
Kramers-Kronig analysis procedure as described in the Supplemental
Material. The results of the Kramers-Kronig analysis are plotted in
Fig.~\ref{overall} in terms of $\sigma_{1}(\omega)$ and the real
part of the dielectric constant,
$\varepsilon^{\prime}(\omega)=1-4\pi\sigma_{2}(\omega)/\omega$.

The first striking result of our investigation is a power-law
behavior of $\sigma_{1}(\omega)$ between approximately 2000 and
8000~cm$^{-1}$: $\sigma_{1}(\omega) \propto \omega^{n}$ with $n =
1.65 \pm 0.05$. This power-law conductivity is basically independent
on temperature.

The straightforward application of Eq.~(\ref{exponent}) yields $z
\simeq 0.6$, i.e., a sublinear dispersion, $E(k) \propto \lvert k
\rvert ^{0.6}$, of the Dirac bands in Cd$_{3}$As$_{2}$. This
dispersion is valid only for the energies above 250~meV ($\sim$
2000~cm$^{-1}$), as the observed power law in $\sigma_{1}(\omega)$
does not extend below this frequency. Also, Eq.~(\ref{exponent})
does not take into account the asymmetry between the valence and
conduction bands, which is present in Cd$_{3}$As$_{2}$
\cite{Liu2014Nat, Jeon2014}. Hence, the obtained dispersion is
basically an effective approximation. Nevertheless, the sublinear
dispersion at high energies is in qualitative agreement with the
dispersion derived from Landau-level spectroscopy \cite{Jeon2014},
with ARPES results \cite{Neupane2014}, as well as with
band-structure calculations \cite{Borisenko2014, Jeon2014}. One
could imagine that the Dirac bands in Cd$_{3}$As$_{2}$ get more
narrow when the Dirac point is approached as depicted by the sketch
in Fig.~\ref{overall}. Eventually, the bands become linear, but we
cannot probe the linear dispersion by optical-conductivity
measurements, because, as will be discussed below, the Fermi level
in our sample is shifted with respect to the Dirac point.

\begin{figure}[b]
\centering
\includegraphics[width=\columnwidth, clip]{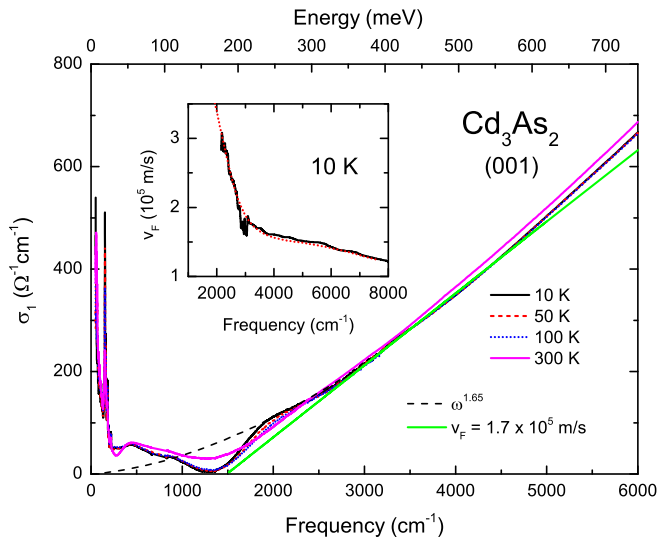}
\caption{(Color online) Optical conductivity from Fig.~\ref{overall}
re-plotted on double-linear scale for $\omega < 6000$~cm$^{-1}$. The
straight green line represents the linear conductivity with $v_{F} =
1.7 \times 10^{5}$~m/s. The inset shows the Fermi velocity
calculated using Eq.~(\ref{v_F}) from $\sigma_{1}(\omega)$ at 10 K.
Red dotted line is a guide to the eye ($v_{F}$ saturates at some
point as $\omega \rightarrow 0$).} \label{mid_freq}
\end{figure}

The dielectric constant (bottom panel of Fig.~\ref{overall}) is
negative at low frequencies and crosses zero at 400~cm$^{-1}$. The
crossing point is independent on temperature and is set by the
screened plasma frequency of the free carriers, $\omega_{\rm
pl}^{\rm scr}$. Very similar values of $\omega_{\rm pl}^{\rm scr}$
have been reported previously \cite{Gelten1980, GeltenPhD,
Houde1986} indicating similar carrier concentrations in naturally
grown samples. At higher frequencies $\varepsilon^{\prime}(\omega)$
takes a positive sign, reaching values up to 17 at 1500 --
2000~cm$^{-1}$.

\section{Discussion}

In the simplest case of symmetric 3D Dirac cones, the slope of the
linear $\sigma_{1}(\omega)$ due to interband contributions is
directly related to the (isotropic) Fermi velocity of Dirac fermions
\cite{Hosur2012,Bacsi2013, Timusk2013}:
\begin{equation}
\sigma_{1}(\omega) = \frac{e^2 N_{W}} {12 h} \frac{\omega} {v_F},
\label{interband}
\end{equation}
where $N_{W}$ is the number of nondegenerate cones and all Dirac
points are considered to be at the Fermi level. If the Fermi level
is not at the Dirac point ($E_{F} \neq 0$), Eq.~(\ref{interband}) is
replaced by \cite{Ashby2014}:
\begin{equation}
\sigma_{1}(\omega) = \frac{e^2 N_{W}} {12 h} \frac{\omega} {v_F}
\theta\left\{\hbar\omega-2E_{F}\right\}, \label{Ashby}
\end{equation}
where $\theta\{x\}$ is the Heaviside step function and any carrier
scattering is ignored.

In Fig.~\ref{mid_freq} we re-plot $\sigma_{1}(\omega)$ on a
double-linear scale as relevant for further considerations. From
Figs.~\ref{overall} and \ref{mid_freq} one can see that the
low-temperature $\sigma_{1}(\omega)$ almost vanishes ($\sim 5$
$\Omega^{-1}$cm$^{-1}$) at around 1300 cm$^{-1}$ (160 meV). The
power-law conductivity starts at $\sim$ 2000 cm$^{-1}$ (250 meV).
Following Ref.~\onlinecite{Ashby2014} and Eq.~(\ref{Ashby}), we
associate this steplike feature in $\sigma_{1}(\omega)$ with the
position of the Fermi level.

The power-law conductivity discussed above can be roughly
approximated by a straight line (the best fit is achieved between
3000 and 4500~cm$^{-1}$). By setting $N_{W} = 4$ (two
spin-degenerate cones) in Eq.~(\ref{interband}), we obtain $v_{F} =
1.7 \times 10^{5}$~m/s. Alternatively, one can estimate $v_{F}$ from
the derivative of the interband conductivity, i.e. from
$\sigma_{1}$($\omega$) at $\omega/2\pi c > 2000$~cm$^{-1}$:
\begin{equation}
v_F(\omega) = \frac{e^2 N_{W}} {12 h} \left(\frac {d\sigma_{1}}
{d\omega} \right) ^{-1}, \label{v_F}
\end{equation}
as it is shown in the inset of Fig.~\ref{mid_freq}. As it can be
seen from the figure, the typical values of $v_{F}$ ($1.2\times
10^{5}$ to $3 \times 10^{5}$~m/s) are somewhat smaller than the
published results, which range from $7.6 \times 10^{5}$~m/s to $1.5
\times 10^{6}$~m/s \cite{Borisenko2014, Liu2014Nat, Neupane2014,
Liang2015, Jeon2014, Cao2015, He2014}. Note however, that we have
evaluated an energy-dependent Fermi velocity: $v_{F}$ strongly
increases as $\omega$ is reduced. In other words, $v_{F}$ might
actually meet the literature values when approaching the Dirac
points. The rise of $v_{F}$ at $\omega \rightarrow 0$ is exactly the
behavior expected for narrowing Dirac bands.

\begin{figure}[b]
\centering
\includegraphics[width=\columnwidth, clip]{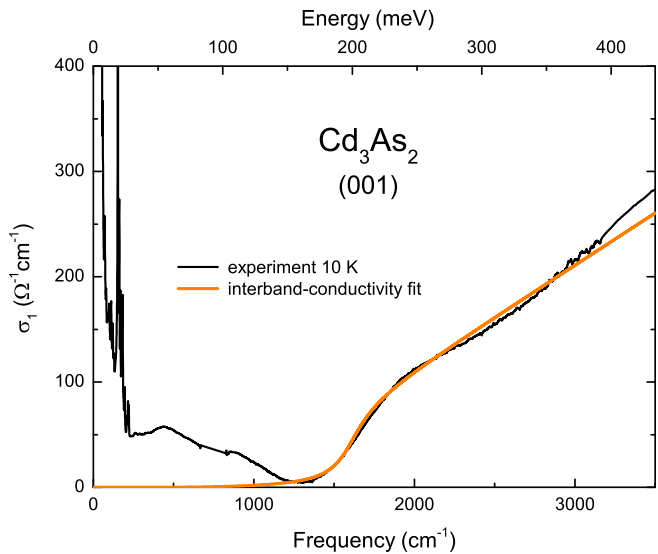}
\caption{(Color online) Optical conductivity of Cd$_{3}$As$_{2}$
measured within the (001) plane at $T=10$~K, plotted together with a
description of the interband portion of $\sigma_{1}(\omega)$ given
by Eqs.~(\ref{sigma_no_imp}) and (\ref{arctangent}). Deviations
between the experiment and theory at higher frequencies are due to
neglecting the deviations from linearity of the Dirac bands in
Eqs.~(\ref{sigma_no_imp}) and (\ref{arctangent}), as discussed in
the text.} \label{self_energy}
\end{figure}

In order to get a more quantitative description of the optical
conductivity in Cd$_{3}$As$_{2}$, one needs to implement a model. So
far, there is no accurate model for the optical conductivity of a 3D
\textit{doped} Dirac semimetal with sublinear band dispersion.
Building such a model goes beyond the present work \cite{Lifshitz}.
Below, we will use a simple model of linear $E(k)$ spectrum with the
purpose of extracting the Fermi-level position and explaining its
broadening. This model will also effectively describe the leveling
off the interband conductivity spectrum at low frequencies.

The model is based on considering self-energy effects in the spirit
of Ref.~\onlinecite{Benfatto2008} and is given in the Supplemental
Material. The real parts of the conduction and valence-band self
energies are approximated by the constant Re$\Sigma_{c} = \Delta$
and Re$\Sigma_{v} = -\Delta$, respectively. We hence obtain for the
interband conductivity
\begin{equation}
\sigma_{1}(\omega) = \frac{e^2 N_{W}} {12 h v_F}
\frac{(\omega-\omega_{g})^{2}} {\omega}
\theta\left\{\hbar\omega-2\max[E_{F},\Delta]\right\},
\label{sigma_no_imp}
\end{equation}
where $\hbar \omega_{g} = 2\Delta$ \cite{Morimoto}. The
conduction-band energies have been pushed up by $\Delta$, while the
valence-band energies are lowered by the same amount. Alternatively,
Eq.~(\ref{sigma_no_imp}) is the interband conductivity of a simple
band structure with $E(k) = \pm [\hbar v_{F} k + \Delta]$, which
could be thought of as a first rough approximation to the case of
the Dirac cones that get narrow as energy approaches the Dirac
point. The Fermi energy is measured from the Dirac point without
self-energy corrections included, and so $E_{F}$ must be larger than
$\Delta$ for finite doping away from charge neutrality. If impurity
scattering cannot be neglected, the Heaviside function can be
replaced by
\begin{equation}
\frac{1}{2}+\frac{1}{\pi}\arctan
\frac{\omega-2\max[E_{F},\Delta]/\hbar}{\gamma}. \label{arctangent}
\end{equation}
where $\gamma$ represents a frequency-independent impurity
scattering rate.

A combination of Eqs.~(\ref{sigma_no_imp}) and (\ref{arctangent}) is
now employed to model the experimental data; the best fit is plotted
in Fig.~\ref{self_energy}. Note, that this model does not include
the intraband conductivity. The best description of the experimental
curve was achieved with $E_{F}/hc = 800$~cm$^{-1}$ ($E_{F} \approx
100$~meV), $\omega_{g}/2\pi c = 450$~cm$^{-1}$, $\gamma/2\pi c =
120$~cm$^{-1}$, and $v_{F} = 2.4 \times 10^{5}$~m/s. The value of
$\omega_{g}$, which is found to be smaller than the Fermi energy,
has to be considered as a fit parameter only. The fit is not
perfect, as the model doesn't include the deviations from the band
linearity discussed above. Nevertheless, the model grasps the main
features of the interband conductivity in Cd$_{3}$As$_{2}$.

The obtained position of the Fermi level (100 meV) seems to be quite
reasonable for our sample, taking into account its carrier
concentration ($6 \times 10^{17}$ cm$^{-3}$) and keeping in mind
that $E_{F} = 200$~meV was reported for a sample with $n_{e} = 2
\times 10^{18}$~cm$^{-3}$ \cite{Jeon2014} and $E_{F} = 286$~meV for
$n_{e} = 1.67 \times 10^{18}$~cm$^{-3}$ \cite{Cao2015}.

\begin{figure}[t]
\centering
\includegraphics[width=\columnwidth, clip]{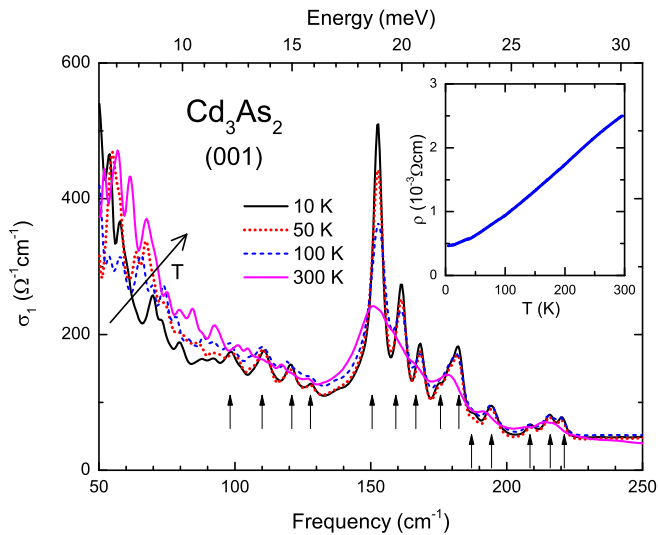}
\caption{(Color online) Low-frequency part of optical conductivity
in Cd$_{3}$As$_{2}$. The vertical arrows indicate the positions of
the phonons. The inset shows dc-resistivity as a function of
temperature measured within the (001) plane.} \label{low_freq}
\end{figure}

Although the scope of this work is the \textit{interband}
conductivity in Cd$_{3}$As$_{2}$, let us briefly discuss the
experimental results at the lowest frequencies measured. The
intraband conductivity is represented by a narrow Drude component
(best seen in Fig.~\ref{low_freq}) and an absorption band of
peculiar shape at $300 - 1300$~cm$^{-1}$ (see Figs.~\ref{mid_freq}
and \ref{self_energy}). The nature of the band might be related to
localization and/or correlation effects. In any case, the presence
of this band makes it impossible to fit the intraband conductivity
for $\omega/2\pi c < 1300$~cm$^{-1}$ with a simple free-electron
Drude term. One can see, however, that the narrow Drude peak is
getting somewhat narrower as $T \rightarrow 0$ due to a modest
decrease of scattering. Let us note that the spectral weight related
to the low-frequency absorbtion of delocalized carriers remains
temperature independent because the screened plasma frequency,
discussed above in connection to the dielectric constant, is
independent of temperature.

In addition to electronic contributions, the low-frequency
conductivity renders a large number of phonon modes marked by arrows
in Fig.~\ref{low_freq}. We can distinguish 14 infrared-active phonon
modes in the frequencies between approximately 100 and
250~cm$^{-1}$. As the temperature is reduced, the phonons become
sharper. More details on the low-frequency conductivity in
Cd$_{3}$As$_{2}$ will be given in a separate paper.

\section{Conclusions}

We found the dc resistivity and the optical conductivity of
[001]-oriented $n$-doped ($n_{e} = 6 \times 10^{17}$~cm$^{-3}$)
Cd$_{3}$As$_{2}$ to be isotropic within the (001) plane. The real
part of the frequency-dependent conductivity follows a power law,
$\sigma_{1}(\omega) \propto \omega^{1.65}$, in a broad frequency
range, 2000 to 8000~cm$^{-1}$. We interpret this behavior as the
manifestation of interband transitions between two Dirac bands with
a sublinear dispersion relation, $E(k) \propto \lvert k \rvert
^{0.6}$. The Fermi velocity falls in the range between $1.2\times
10^{5}$ and $3 \times 10^{5}$~m/s, depending on the distance from
the Dirac points.

At 1300~cm$^{-1}$ (160~meV), we found a diminishing conductivity,
consistent with observations of an ``optical gap'', made in the
1960s \cite{Turner1961}. However, we hesitate to follow the
traditional interpretation of this feature and to straightforwardly
relate it to a gap in the density of states. Applying recent models
for the optical response of Dirac/Weyl semimetals, we instead relate
this feature to the Fermi level, which is positioned around
$100$~meV above the Dirac points, and which is consistent with the
carrier concentration.

\section{Acknowledgements}

\begin{acknowledgments}
Ruud Hendrikx at the Department of Materials Science and Engineering
of the Delft University of Technology is acknowledged for the x-ray
analysis. We are grateful to Gabriele Untereiner for sample
preparation and other technical assistance. Sample-characterization
work of Hadj M. Bania is highly appreciated. We thank Hongbin Zhang,
Sergue\"{i} Tchoumakov, Marcello Civelli, Robert Triebl, and Gernot
Kraberger for their efforts in $ab-initio$ calculations of the
Cd$_{3}$As$_{2}$ optical conductivity, for sharing their results
with us, and for many useful conversations. A. A. N. is grateful to
Prof. Ernest Arushanov for his interest in this work and useful
discussions. This work was supported by the Deutsche
Forschungsgemeinschaft (DFG).
\end{acknowledgments}

\section{Supplemental material}

\subsection{Sample growth}

Single crystals of Cd$_{3}$As$_{2}$ have been grown by vapor
transport from material previously synthesized in argon flow with
the flow rate of a few cm$^{3}$/min \cite{Nateprov2015}. The
temperatures in the evaporation and growth zones are 520 $^{\circ}$C
and 480 $^{\circ}$C, respectively. By subsequent annealing at room
temperature the electron mobility is enhanced while at the same time
the electron concentration decreases \cite{Rambo1979}. Well-defined
crystals with facets measured a few millimeters in each direction
are harvested after 24 hours of the growth process.

The lattice parameters have been obtained by x-ray diffraction: $a =
1.267$~nm and $c = 2.548$~nm, in good agrement with recent
measurements reporting a tetragonal unit cell with only light
deviations from a cubic structure \cite{Ali2014}.

\subsection{Kramers-Kronig analysis}

For Kramers-Kronig analysis, extrapolations to zero frequency have
been made using the Hagen-Rubens relation in accordance with the
temperature-dependent dc resistivity measurements (Fig. 5 of the
article). For the temperature-independent extrapolation to $\omega
\rightarrow \infty$, we utilized published reflectivity data,
obtained in ultra-high vacuum for up to 170\,000~cm$^{-1}$ (21 eV)
\cite{Zivitz}, applying the procedure proposed by Tanner
\cite{Tanner2015}, which utilizes x-ray atomic scattering functions
for calculating the optical response for frequencies from 21 to
30\,000~eV and the free-electron $R \propto \omega^{-4}$
extrapolation for frequencies above 30~keV.

In order to check the robustness of our Kramers-Kronig analysis, we
replaced the reflectivity data of Ref.~[\onlinecite{Zivitz}] by
three other data sets available in the literature for the frequency
range from approximately 1 to 5 eV. Namely, we used: (i) the
polycrystalline data from Ref.~[\onlinecite{Karnicka}]; (ii) the
single-crystal data from the same reference; and (iii) the most
recent single-crystal data from Kozlov {\it et al.} \cite{Kozlov}.
All these data sets, as well as the data of
Ref.~[\onlinecite{Zivitz}], show rather similar overall reflectivity
with two strong peaks at approximately 1.8 and 4~eV; the peak at
1.8~eV shows up in our data as well. We have found that at
frequencies below 1.2~eV (10\,000~cm$^{-1}$) all these different
extrapolations affect the outcome of the Kramers-Kronig analysis
only marginally. However, it is essential not to neglect the strong
peak in reflection at around 4~eV. If we ignore the 4-eV peak in our
Kramers-Kronig analysis and use instead the free-electron
($\omega^{-4}$) extrapolation right from the highest data point, the
exponent of the power-law dependence of the optical conductivity
between 2000 and 8000~cm$^{-1}$, discussed in the article, will be
reduced from 1.65 to 1.2.

\subsection{Interband optical conductivity of linearly dispersed
bands with self-energy effects}

For a Dirac/Weyl semimetal in the pure limit with no impurity
scattering, the dynamic longitudinal optical conductivity associated
with a single Weyl point can be expressed as \cite{Ashby2014}:
\begin{align}
\nonumber \sigma_{1}(\omega) = \frac{e^2} {3 \pi v_F \omega}
\int_{E_{F}-\omega}^{E_{F}}{d\omega'}\times\\\int_{0}^{\infty}{\epsilon^{2}
d\epsilon [\delta(\omega'-\epsilon)\delta(\omega'+\omega+\epsilon) +
\delta(\omega'+\epsilon)\delta(\omega'+\omega-\epsilon)]}
\label{sg1}
\end{align}
(we adopt $\hbar = 1$ for calculations here). In the presence of
self-energy effects, the $\delta$-functions in Eq.~(\ref{sg1}) are
to be replaced by the corresponding conduction ($c$) and valence
($v$) band spectral functions:
\begin{align}
A_{c,v}(\epsilon,\omega') =
\frac{1}{\pi}\frac{-\textrm{Im}\Sigma_{c,v}(\omega')}
{\left(\omega'-\textrm{Re}\Sigma_{c,v}(\omega')\mp\epsilon\right)^2+\left(\textrm{Im}\Sigma_{c,v}(\omega')\right)^2}.
\label{sg2}
\end{align}

For our purposes, it will be sufficient to take the imaginary part
of the self energy to be infinitesimal small and the real part of
the self energy to be some constant with $\Sigma_{v} = -\Sigma_{c}$,
so that the conduction band is translated up by $\Sigma_{c} =
\Delta$ and the valence band down by $\Sigma_{v} = -\Delta$. The
real part of the conductivity then takes the form:
\begin{align}
\nonumber \sigma_{1}(\omega) = \frac{e^2} {3 \pi v_F \omega}
\int_{E_{F}-\omega}^{E_{F}}{d\omega'}\\ \nonumber \times
\int_{0}^{\infty} \epsilon^{2} d\epsilon
[\delta(\omega'-\Delta-\epsilon)\delta(\omega'+\omega+\Delta+\epsilon)\\
+\delta(\omega'+\Delta+\epsilon)\delta(\omega'+\omega-\Delta-\epsilon)],
\label{sg3}
\end{align}
which works out to be
\begin{equation}
\sigma_{1}(\omega) = \frac{e^2} {24 \pi v_F}
\frac{(\omega-2\Delta)}{\omega}^{2}\theta(\omega-2\max[E_{F}
\Delta]). \label{sg4}
\end{equation}
Note, that $2\Delta$ is the optical gap and $\sigma_{1}(\omega)$ is
zero for all photon energies below $2\Delta$ when $E_{F}=0$. If the
system is doped ($E_{F}>\Delta$), then the lower cut-off in the
conductivity is $2E_{F}$ rather than $2\Delta$.


\begin{thebibliography}{99}
\bibitem{Lifshitz} The electron-hole asymmetry of the Dirac bands as well as the
influence of the Lifshitz-transition points, which exact position is
debatable, have to be taken into account in a complete
optical-conductivity model for Cd$_{3}$As$_{2}$.
\end{thebibliography}
\end{document}